\documentclass[sigconf,nonacm]{acmart}
\usepackage{colortbl}

\AtBeginDocument{%
  }

\begin{document}

\title{Using LLMs to Mimic the Conversational Dynamics of Reddit Communities}



\author{Vedaant Jain}
\email{vvjain3@illinois.edu}
\affiliation{%
  \institution{University of Illinois Urbana-Champaign}
  \city{Urbana}
  \state{IL}
  \country{USA}
}
\authornote{These authors contributed equally to the paper.}

\author{Yoshee Jain}
\email{yosheej2@illinois.edu}
\affiliation{%
  \institution{University of Illinois Urbana-Champaign}
  \city{Urbana}
  \state{IL}
  \country{USA}
}
\authornotemark[1]

\author{Ishq Gupta}
\email{ig8@illinois.edu}
\affiliation{%
  \institution{University of Illinois Urbana-Champaign}
  \city{Urbana}
  \state{IL}
  \country{USA}
}

\author{Aditi Shrivastava}
\email{aditi12@illinois.edu}
\affiliation{%
  \institution{University of Illinois Urbana-Champaign}
  \city{Urbana}
  \state{IL}
  \country{USA}
}

\author{Koustuv Saha}
\email{ksaha2@illinois.edu}
\affiliation{%
  \institution{University of Illinois Urbana-Champaign}
  \city{Urbana}
  \state{IL}
  \country{USA}
}

\author{Eshwar Chandrasekharan}
\email{eshwar@illinois.edu}
\affiliation{%
  \institution{University of Illinois Urbana-Champaign}
  \city{Urbana}
  \state{IL}
  \country{USA}
}

\renewcommand{\shortauthors}{Jain et al.}

\ccsdesc[500]{Human-centered computing~Collaborative and social computing}

\keywords{Social computing, large language models, prosocial behavior}


\begin{abstract}
Online communities face a constant battle against toxicity and misinformation. While human moderators struggle to keep pace with the volume of content, LLMs offer a promising solution for automatically generating constructive responses and shaping online interactions. This paper preliminarily investigates if LLMs can mimic the communication styles of Reddit users using their comment history as context. We evaluate two prompting approaches: predicting a target comment and filling in masked comments. We find that LLMs outperform expectations at replicating comment structure and formality, but struggle to accurately capture nuanced emotions, e.g. understating joy and overstating anger. These findings highlight a promising direction for LLMs in guiding online conversations towards prosociality influencing emergent communication patterns and norms within the community. The results of our study inspire future work with more rigorous methods of evaluation to explore the LLMs' effectiveness across diverse online communities to better understand their broader societal impact.
\end{abstract}

\maketitle

\section{Introduction and Background} 

The rise of social media necessitates fostering positive and constructive online environments. Moderators on platforms like Reddit face increasing challenges in maintaining healthy communities, grappling with a deluge of interactions and user behavior. Large Language Models (LLMs) offer a promising avenue for addressing this challenge. \citet{ziems2024large} highlight LLMs' capabilties on zero-shot annotation and creative tasks suggesting their viability as tools to augment human effort. While previous work has demonstrated the potential of LLMs in simulating social media dialogue, the focus has often been on generating synthetic content based on general user descriptions \cite{park2022social} or simulating specific conversational roles \cite{abbasiantaeb2023let}. LLMs have also proven valuable in tasks such as evaluation mimicking \cite{Aiyappa_2023, guo2023close, pegoraro2023chatgpt, qin2023chatgpt} and document generation \cite{askari2023generating}. However, these approaches lack the user-specific context necessary for accurately capturing individual communication nuances.

This paper explores the potential of LLMs to accurately simulate human social behavior in realistic conversational settings by investigating if LLMs can effectively mimic the communication patterns of users within Reddit communities when provided with their past interactions as contextual grounding. Specifically, we address the research question:
\\
\textbf{RQ: How well do LLMs mimic the semantic structure and sentiment of specific users given partial history?}, 

Influenced by findings from \cite{friedman2023leveraging} that demonstrate that LLMs can effectively leverage user-specific data to personalize outputs in conversational recommender systems, we hypothesize that incorporating prior user activity as input for LLMs enables the LLM to develop a comprehensive understanding of individual user characteristics, leading to more accurate and nuanced simulations.

This proposal has the potential to revolutionize online moderation by enabling the development of automated systems that can mitigate "anti-social" behavior by generating strategic responses aimed at guiding conversations towards neutrality or even prosociality \cite{batson2003altruism_prosocial_behavior}. Moreover, effective LLM-powered synthetic data generators could provide researchers with access to high-quality, ethically-sourced data, overcoming prior concerns surrounding privacy, data volume, and consent, accelerating advancements in research.

\section{Methodology}

To begin our preliminary analysis, we collect data from the r/science subreddit using the publicly available Reddit API: PRAW \cite{praw}. r/science offers a highly moderated environment focusing on factual discourse, providing a controlled and predictable setting for evaluating LLM performance. This environment simplifies the modeling task for LLMs, allowing us to focus on the core research question shadowing the complexities introduced by diverse and unmoderated content.

We scraped a comprehensive dataset, denoted as $\mathcal{D}$, consisting of posts ($n = 12000$), comments, and associated metadata (author information, post and comment timestamps, and the hierarchical structure of comments within threads). The data collection process utilized the "new," "hot," and "controversial" features of the API to capture a diverse range of discussions and user engagement patterns. To enhance the LLM's ability to learn user-specific writing styles, we sub-sampled $\mathcal{D}$, focusing on comments from authors with a minimum of eight prior comments. Additionally, we only considered comments with at least five parent comments to ensure sufficient contextual information for accurate prediction. 

Within $\mathcal{D}$, each post \( P_i \) is associated with a number \( T_{i_{\text{num}}} \) of threads. A thread ($T_{ij}$) represents a hierarchical chain of comments. Each thread has an \textbf{initial comment}; the first comment in the thread ($C_{ij1}$) which is a direct reply to the original post ($P_i$). Each thread also has \textbf{subsequent comments} where each subsequent comment ($C_{ijk}$, $k > 1$) is a direct reply to its parent\footnote{Note: Definition of parent: If a comment $c$ is the parent of comment $d$ then $d$ is a direct reply to $c$.\label{fn:parent}} comment ($C_{ij(k-1)}$) within the thread.

Our experiments utilized the Gemini 1.0-pro model\cite{geminiteam2024gemini} with a temperature setting of 1.0. We explored two distinct settings for modeling comments using LLMs: the \textit{predict setting} and \textit{masked fill-in-the-blank setting}.

\begin{figure*}
    \centering
    \begin{minipage}{0.49\textwidth}
        \includegraphics[width=\linewidth]{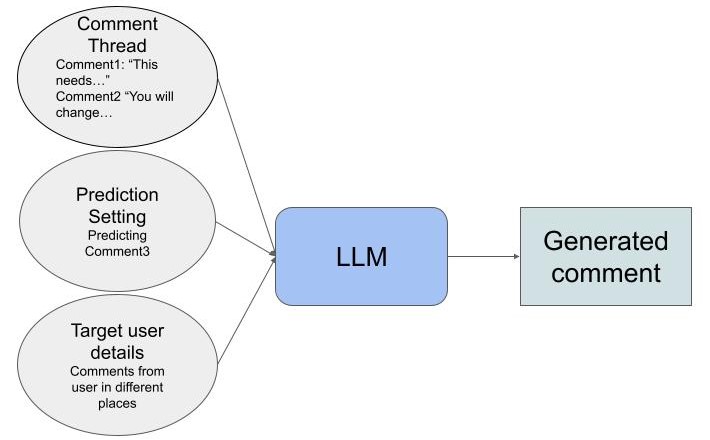}
        \caption{`Fill' Prompt Pipeline}
        \label{fig:left_pic}
    \end{minipage}\hfill
    \begin{minipage}{0.49\textwidth}
        \includegraphics[width=\linewidth]{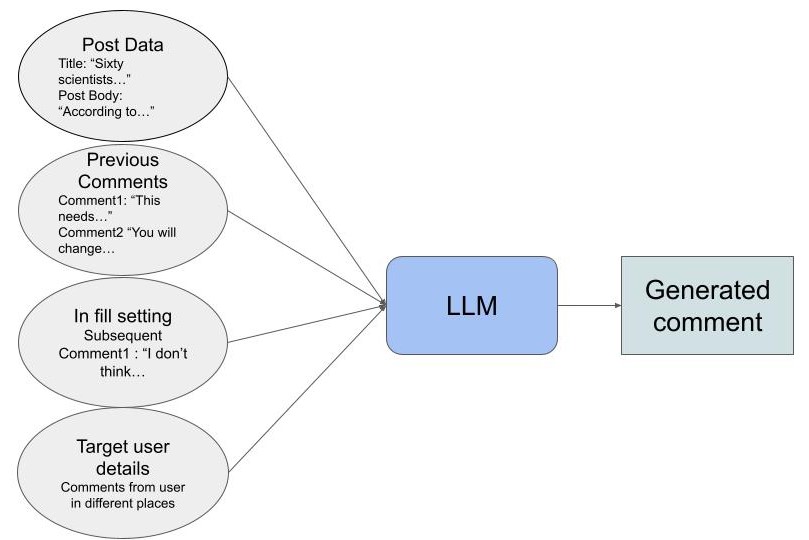}
        \caption{`Predict' Prompt Pipeline Illustration}
        \label{fig:right_pic}
    \end{minipage}
\end{figure*}

\subsection{Predict Setting}

This setting focuses on predicting a target comment ($C_{ijk}$) based on its preceding conversational context within the thread. As depicted in Figure~\ref{fig:left_pic}, the input to the LLM consists of the \textit{ancestral comment history}, \textit{post content}, and \textit{user history}. The \textit{ancestral comment history} are all comments preceding the target comment $C_{ijl}$ where $l<k$. The post Content is defined as the title and body of the original post ($P_i$). The user history refers to the selection of previous comments made by the target comment's author within the r/science subreddit.

These elements are combined into a single prompt, instructing the LLM to \texttt{analyze a target user's writing style from their comments in other threads/posts.} We instruct the LLM to \texttt{mimic the style of the user to respond to the given Reddit thread, addressing the last comment while staying relevant to the post.}

\subsection{Masked-Fill-in-the-blank Setting}

In this setting, the LLM is tasked with predicting a comment ($C_{ijk}$) given both its ancestral and successor comments within the thread ($C_{ijl}$ for $l \neq k$) (illustrated in Figure~\ref{fig:right_pic}). Similar to the predict setting, the LLM receives the post content, user history, and surrounding comments as context. The LLM is asked to
\texttt{create a comment that seamlessly fits into the existing Reddit conversation thread by mimicking the writing style of the specified Reddit user.}

\section{Evaluation}

To understand the performance of Gemini in imitating Reddit users, we evaluate the LLM-generated responses in comparison to actual data collected from Reddit (we refer to this data as ``ground truth''). Our evaluation criteria for this analysis (keeping the ``ground truth'' comments as a baseline) was two-fold: style similarity and content similarity.

For quantifying style similarity, we measured the following using publicly available pre-trained classifiers: formality of the text \cite{formality_classifier}, an understanding of whether the response is a question or statement \cite{question_statement_classifier}, and syntactic similarity \cite{natural_language_ai}. A similar methodology follows for assessing content similarity: parallelism between underlying meanings \cite{sentence_similarity_classifier}, emotional analysis \cite{sentiment_analysis_classfier}, and sentiment analysis \cite{natural_language_ai}. In addition, we also examine how well the LLM imitates different types of users and emotions by measuring the emulated user's comments that were pre-classified into distinct emotion categories (joy, sadness, anger, etc.).

To this end, we define two sets of distributions for each feature, $X_{f_i\text{LLM}}$ and $X_{f_i\text{Truth}}$, where $f_i$ represents a specific feature extracted from each comment in the dataset. These distributions encompass the values corresponding to each feature from the data points. We hypothesize that these distributions generally conform to a normal distribution, which is validated through visual examination of histograms and quantile-quantile plots.

\subsection{Comparison of Distributions}
To compare the ``ground truth'' data with the LLM-generated results, we define $X_{f_i\text{diff}} = X_{f_i\text{Truth}} - X_{f_i\text{LLM}}$. As linear combinations of normal distributions also yield a normal distribution, $X_{f_i\text{diff}}$ is assumed to be normally distributed. 

Subsequently, we perform hypothesis testing using the $t$-test \cite{student1908probable} on $X_{f_i\text{diff}}$ to examine if $H_0: \mu = 0$; the mean difference is zero, indicating no bias. Moreover, we explore $H_0: \mu <= 0$ where the mean difference is positive suggesting that LLM estimates are systematically lower than the truth. We also evaluate its counterpart: $H_0: \mu >= 0$. In all these hypotheses, $\mu$ represents the mean of the $X_{f_i\text{diff}}$ distribution.

\subsection{Categorization of Continuous Features}

To address the inherent interdependency of probabilistic outputs from softmax functions used in the classification models, we transform continuous output scores into categorical variables. Directly comparing these continuous scores can be misleading due to their inherent correlation (summing to one). For instance, a high joy score (e.g., 0.95) automatically dictates a low sadness score (e.g., 0.05), potentially misrepresenting the presence of sadness. To avoid incorporating values that are not meaningful into our analysis, we categorize the scores; a score $> 0.3$ is assigned a value of 1 (significant presence of the feature) and a score of ($<= 0.3$) is assigned a value of 0 (minimal influence). This threshold conveys clear stylistic or emotional dominance, enhancing interpretability and analysis robustness to allow us to focus on substantial feature expressions.

We utilize two methods for assessing feature-wise accuracy, namely \textit{grouped average} and \textit{filtered average}. \textit{Grouped average} represents the mean accuracy across all considered features from a model whereas \textit{filtered average} exclusively considers features classified as 1 (significantly present) in the ground truth comment, reflecting the expectation that the LLM should replicate these dominant features.

\section{Results}

\begin{table*}[ht]
\small
\centering
\caption{Consolidated T-test results for both tasks across various features showing statistically significant results with an * for each hypothesis. An absence of data generated for the predict setting for certain metrics due to computational limitations has been highlighted with a color of gray.}
\label{tab:consolidated-fill-task-results}
\begin{tabular}{@{}l|ccc|ccc@{}}
\toprule
\multicolumn{1}{c}{\textbf{Feature}} & \multicolumn{3}{c}{\textbf{Masked Fill-in-the-Blank}} & \multicolumn{3}{c}{\textbf{Predict}} \\
 & \textbf{$\mu = 0$} & \textbf{$\mu <= 0$} & \textbf{$\mu >= 0$} & \textbf{$\mu = 0$} & \textbf{$\mu <= 0$} & \textbf{$\mu >= 0$} \\
\midrule
Sadness & * &  & * & * & * & \\
Joy & * & & * & \multicolumn{3}{c}{\cellcolor[gray]{0.9}} \\
Love & * &  & * & \multicolumn{3}{c}{\cellcolor[gray]{0.9}} \\
Anger & * & * &  & \multicolumn{3}{c}{\cellcolor[gray]{0.9}} \\
Fear & * & & * & \multicolumn{3}{c}{\cellcolor[gray]{0.9}} \\
Surprise & * & & * & * &  & * \\
Formal & * &  & * & * & & * \\
Informal & * & * &  & * & * & \\
Statement & * &  & * & \multicolumn{3}{c}{\cellcolor[gray]{0.9}} \\
Question & * & * &  & \multicolumn{3}{c}{\cellcolor[gray]{0.9}} \\
\end{tabular}
\end{table*}

\begin{table*}[ht]
\small
\centering
\caption{Categorical accuracy results for both tasks displaying individual feature accuracies and their grouped averages.}
\label{tab:categorical-accuracy-results}
\begin{tabular}{@{}ll|r|r@{}}
\toprule
\textbf{Feature Group} & \textbf{Feature Pair} & \textbf{Masked Fill-in-the-Blank Setting} & \textbf{Predict Setting} \\
\midrule
\multicolumn{3}{c}{\textit{Emotional and Stylistic Features}} \\
\midrule
Emotions & Sadness & 0.865 & 0.887\\
 & Joy & 0.653 & 0.647 \\
 & Love & 0.991 & 0.993 \\
 & Anger & 0.538 & 0.540 \\
 & Fear & 0.873 & 0.874 \\
 & Surprise & 0.982 & 0.989 \\
\textbf{Grouped Average} & \textbf{All Emotions} & \textbf{0.817} & \textbf{0.822} \\
\textbf{Filtered Average} & \textbf{Features $>$ 0} & \textbf{0.593} & \textbf{0.611} \\

\midrule
\multicolumn{3}{c}{\textit{Formality Features}} \\
\midrule
Formality & Formal & 0.884 & 0.856 \\
 & Informal & 0.743 & 0.496 \\
\textbf{Grouped Average} & \textbf{All Formality} & \textbf{0.681} & \textbf{0.676} \\
\textbf{Filtered Average} & \textbf{All Formality} & \textbf{0.743} & \textbf{0.741} \\

\midrule
\multicolumn{3}{c}{\textit{Comment Structure Features}} \\
\midrule
Labels & Statement & 0.862 & 0.858 \\
 & Question & 0.937 & 0.941 \\
\textbf{Grouped Average} & \textbf{Comment Structure} & \textbf{0.900} & \textbf{0.899} \\
\textbf{Filtered Average} & \textbf{Comment Structure} & \textbf{0.912} & \textbf{0.909} \\
\bottomrule
\end{tabular}
\end{table*}

Generally, when generating a comment after being fed a ‘masked fill-in-the-blank’ style prompt, the LLM is able to able to understand and replicate the emotions of the comment somewhat accurately (grouped average accuracy of 0.817 and an accuracy score of 0.593 when filtered for the significant emotions) as seen in Table~\ref{tab:categorical-accuracy-results}. In particular, it tends to be more accurate when looking at emotions like Sadness (accuracy score of 0.865), Love (0.991), Fear (0.873) and Surprise (0.982).

When analyzing emotions with a comparatively lower accuracy score in Table~\ref{tab:categorical-accuracy-results}, we see that, according to Table~\ref{tab:consolidated-fill-task-results}, they tend to consistently differ from the comment attempting to be imitated. In particular, we see that Gemeni tends to produce comments that have less joy and increased anger when compared to the comments that are being emulated, with $t$-statistic and p-value of -29.79 and 1.00 ($H_0: \mu <= 0$) for joy and 42.96 and 1.00 ($H_0: \mu >= 0$) for anger. 

For comments generated after being fed a ‘predict’ style prompt, we see that it follows the same patterns as ‘fill-in-the-blank’, with higher accuracy for emotions like sadness (0.887), love (0.993), fear (0.874) and surprise (0.989) and lower accuracy for emotions joy and anger as seen in Table~\ref{tab:categorical-accuracy-results}. Comments generated by ‘predict’ prompts also tend to indicate lower joy and higher anger than the comments they are attempting to emulate with joy having a high p-value for $H_0: \mu <= 0$ (1.00) and anger having a high p-value for $H_0: \mu >= 0$ (1.00) (see Table~\ref{tab:consolidated-fill-task-results}).This indicates that regardless of prompt style, the LLM tends to generate comments that are more angry and less joyous than an intended user for a similar situation.

In terms of emulating user formality, Gemini tends to be more accurate, with an accuracy score for the filtered average at 0.743 for comments generated with a ‘fill’ prompt and a score of 0.741 for comments generated with a ‘predict’ prompt (indicated in Table~\ref{tab:categorical-accuracy-results}). While for both kinds of comments, Gemini tends to be more accurate, it still tends to skew towards a more formal tone than the user, as seen in the $t$-statistic and corresponding p-value scores (1.00 for both) in Table~\ref{tab:consolidated-fill-task-results}. With the general informality of Reddit users, despite the relative accuracy of formality for the given context, a higher level of formality in the generated comments may unveil the LLM diguised comments to other users. 

Like formality, Gemini tends to be more accurate for sentence structure (filtered average accuracy = 0.912 for ‘fill-in-the-blank’ and 0.909 for ‘predict’). As seen in Table~\ref{tab:categorical-accuracy-results} and Table~\ref{tab:consolidated-fill-task-results}, we can also see that Gemini also tends to skew the comment structure towards a statement as opposed to a question. 

Overall, Gemini seems to either understate or overstate certain emotions - in particular, understating joy and overstating anger. In an environment like r/Science, where a pro-social comment would be characterized as being less inflammatory (thus less angry) having a comment that is more angry and less joyous may not help steer the conversation in an intended direction. However, in terms of comment structure and formality, Gemini tends to be more accurate. So, while LLMs like Gemini may be good at emulating user style, it needs to get better at assessing the emotional intensity of a given comment to generate comments that fit into the context of threads on r/Science.

\section{Conclusion}

Our findings indicate a promising yet nuanced landscape for leveraging LLMs in online moderation through synthetic comment generation. Gemini demonstrates a strong capacity to capture stylistic elements of user communication, particularly regarding formality and comment structure. This suggests potential for generating responses that seamlessly integrate into existing conversations, a crucial factor for influencing online discourse organically.
However, the discrepancy in accurately replicating emotional nuances presents a critical area for improvement. The observed tendency to understate joy and overstate anger, regardless of prompting style, underscores a key challenge: LLMs may struggle to grasp the subtle emotional undertones crucial for navigating sensitive online discussions. In a community like r/science, where objectivity and respectful debate are paramount, this could inadvertently escalate tensions rather than fostering a more neutral or prosocial environment. While our preliminary analysis showcases the potential of LLMs for content moderation by imitating user style based on comment history, it also highlights the need for refining their ability to accurately reflect and navigate the complexities of human emotion in online communication.
This study represents an initial exploration into a promising new direction for online moderation using LLM-generated content. Future work should focus on expanding these findings through larger-scale studies incorporating diverse datasets from multiple subreddits and encompassing a wider range of state-of-the-art LLMs for more rigorous evaluations. Additionally, employing more robust analytical techniques, such as ANOVA, would enable a deeper understanding of the interplay between various factors influencing LLM performance. This multifaceted approach is crucial for developing a comprehensive understanding of LLMs' capabilities and limitations in online moderation, paving the way for their effective deployment in augmenting human efforts to foster healthier and more positive online communities.

\bibliographystyle{ACM-Reference-Format}
\bibliography{sample-base}

\end{document}